\documentclass[12pt]{iopart}
\usepackage{epsfig}
\usepackage{graphicx}
\begin{document}

\title{Dislocation interactions mediated by grain boundaries}

\author{Paolo Moretti$^{1,2}$, Lasse Laurson$^{1}$, Mikko J. Alava$^{1}$}

\address{$^{1*}$ Laboratory of Physics, Helsinki University of
Technology, FIN-02015 HUT}
\address{$^{2*}$ Departament de F\'\i sica Fonamental, Facultat
de F\'isica, Universitat de Barcelona, Marti i Franqu\`es 1, E-08028
Barcelona, Spain} 
\ead{paolo.moretti@ub.edu, lla@fyslab.hut.fi, mja@fyslab.hut.fi}

\begin{abstract}
The dynamics of dislocation assemblies in deforming crystals indicate 
the emergence of collective phenomena, intermittent fluctuations and 
strain avalanches. In polycrystalline materials, the understanding 
of plastic deformation mechanisms depends on grasping the role of 
grain boundaries on dislocation motion. Here the interaction of 
dislocations and elastic, low angle grain boundaries is studied in 
the framework of a discrete dislocation representation. We allow grain 
boundaries to deform under the effect of dislocation stress fields and
compare the effect of such a perturbation to the case of rigid grain 
boudaries. We are able to determine, both analytically and numerically, 
corrections to dislocation stress fields acting on neighboring grains, 
as mediated by grain boundary deformation. Finally, we discuss
conclusions and consequences for the avalanche statistics, as observed in
polycrystalline samples.
\end{abstract}

\pacs{62.20.Mk,05.40.-a, 81.40Np}

\maketitle

\date{\today}

\section{Introduction}

Recent advances in experimental techniques have allowed the
redefinition of the once supposedly well-understood problem of
crystal plasticity. In spite of the traditional picture of a smooth
fluid-like motion, acoustic emission (AE) measurements on deforming
ice crystals \cite{WEI-97,MIG-01} as well as more recent compression
tests on metallic specimens \cite{UCH-04,DIM-06} have during the
last decade made it clear that plastic deformation of crystalline
materials proceeds as a heterogeneous and intermittent adjustment to
applied external drives. Such intermittence in plastic deformation
had been for long observed \cite{TIN-73,POT-83} and dismissed as
considered unworthy of serious consideration.  The recent
discoveries challenge this picture since that heterogeneity appears
to take over through the emergence of a burst-like scale-free
behavior. Plastic deformation is found to develop through strain
avalanches, characterized by power law distributed sizes and energy
emissions. Slip motion is found to follow fractal patterns
\cite{WEI-03} and deformed specimens exhibit self-affine surfaces
\cite{ZAI-04}.

The observation of scale invariance suggested that fluctuations
should not be imputed to the discrete nature of the deforming
lattices. Instead, the collective behavior of crystal
dislocation was held responsible for the emergence of scale-free
patterns, in a close-to-critical behavior fashion. Such an
observation triggered a number of numerical
\cite{MIG-01,RIC-05,CSI-07} and, whenever feasible, analytical
\cite{KOS-04,ZAI-05} studies so that the validity and the limits
of applicability of this picture could be assessed.

At present, there is evidence that the notion of scale invariance in
crystal plasticity as a consequence of a close-to-critical state is
consistent with observation in single and multi-slip geometries,
over a variety of materials, hardening coefficients and loading
conditions. However, the general understanding of these processes is
still chiefly phenomenological and several aspects of such
phenomenology are still obscure.  In particular, it is still not
entirely clear to which extent such considerations should be
extended to the case of polycrystalline geometries.

Grain structure has always been supposed to play a central role in
crystal yielding \cite{HAL-51,PET-53}, and has attracted considerable
attention also more recently \cite{ZHA-04,YAM-02,SHE-98,SHA-04,LEO-07}. 
The so called Hall-Petch law,
which relates the yield stress of a polycrystal to its average grain
size, has been recently reformulated in the light of collective
dislocation dynamics models \cite{LOU-06} and its breakdown for
nanometer sized grains explained as a result of the failure of those
systems to initiate or sustain such a process.

These observations pose the general question on how strain avalanches
spread through grain boundaries (GB). The most simplistic approach
consists in assuming that the elastic interactions of two
dislocations on two sides of a GB are entirely screened by the GB.
Such a crude assumption holds only if the different slip geometries
of neighboring grains allow that. In general, however, nothing
prevents dislocation stresses to act on neighboring grains, at least
in a dampened form due to geometrical incompatibilities.

A reasonably balanced approach consists in postulating that plastic
flow (i.e. dislocation dynamics) within a certain grain builds up
back stresses due to incompatibility of orientations and slip
configurations with the neighboring grains. As soon as the
polycrystal reaches the yield point, the accumulated back stresses
are redistributed within the neighboring grains. Plastic yield is
thus accompanied by a relaxation process by which dislocations
effectively interact through grain boundaries by transferring
incompatibility stresses.

The advantage of such an approach is that its results indirectly
account for experimental observation. By looking at cumulative
probability distributions of emission amplitudes in AE experiments
\cite{RIC-05}, for example, creeping ice crystals exhibit a
power-law decay with a rather universal exponent $\beta$.
Polycrystalline samples still retain the power-law behavior, however
with a lower (and possibly non-universal) exponent and, as expected,
a large size cut-off due to finite grain sizes. A natural
explanation of the cut-off is that to some extent strain avalanches
are confined within grains. However, explorations of avalanche
models \cite{RIC-05} seem to suggest that the decrease of the
exponent $\beta$ is possible only if excess stresses within one
grain can eventually trigger dislocation avalanches in the
neighboring ones.

Dislocation interactions through grain boundaries are thus essential
to explain plastic flow in polycrystals. The idea of incompatibility
stresses acting on neighboring grains certainly captures part of the
physics of the problem but it leaves several questions unanswered,
by virtue of its generality. For instance, the exact mechanism by
which stresses are relaxed is not clear, as well as not much can be
said about the typical range of stress rearrangements on a
quantitative basis.

In this paper, we perform an analytical and numerical study of
dislocation interactions through grain boundaries, in the light of
the above considerations. In order to make the problem treatable, we
consider a relatively simple GB model, that of a low-angle grain
boundary, or dislocation wall, in a two-dimensional geometry. This
serves as a starting point to capture the relevant physics of the
problem and expose it in a conveniently simple way. In passing, we
also recall that dislocation walls are frequently encountered in
deforming crystals as well as in a variety of similar systems such
as colloidal crystals \cite{LIN-05} and flux-line lattices in
Type~II superconductors \cite{MOR-05}.

In the following we discuss how stresses generated by dislocations
deform low-angle grain boundaries and how this affects in return the
GB stress field perceived by dislocations in their proximity.
Theoretical predictions are compared with numerical results. We find
that this perturbative effect accounts for a screening effect for
stress deforming the grain boundaries. The typical strength of this
screening proves inversely proportional to the grain size and the
perturbation itself is found to be short-ranged and exponentially
suppressed beyond a distance that again scales with the grain size.
The analytical and numerical computations are in the next section,
while Section \ref{discussion} is devoted to discussion and conclusions.

\section{GB-dislocation interaction}

\begin{figure}[!h]
\begin{center}
\includegraphics[width = 10cm]{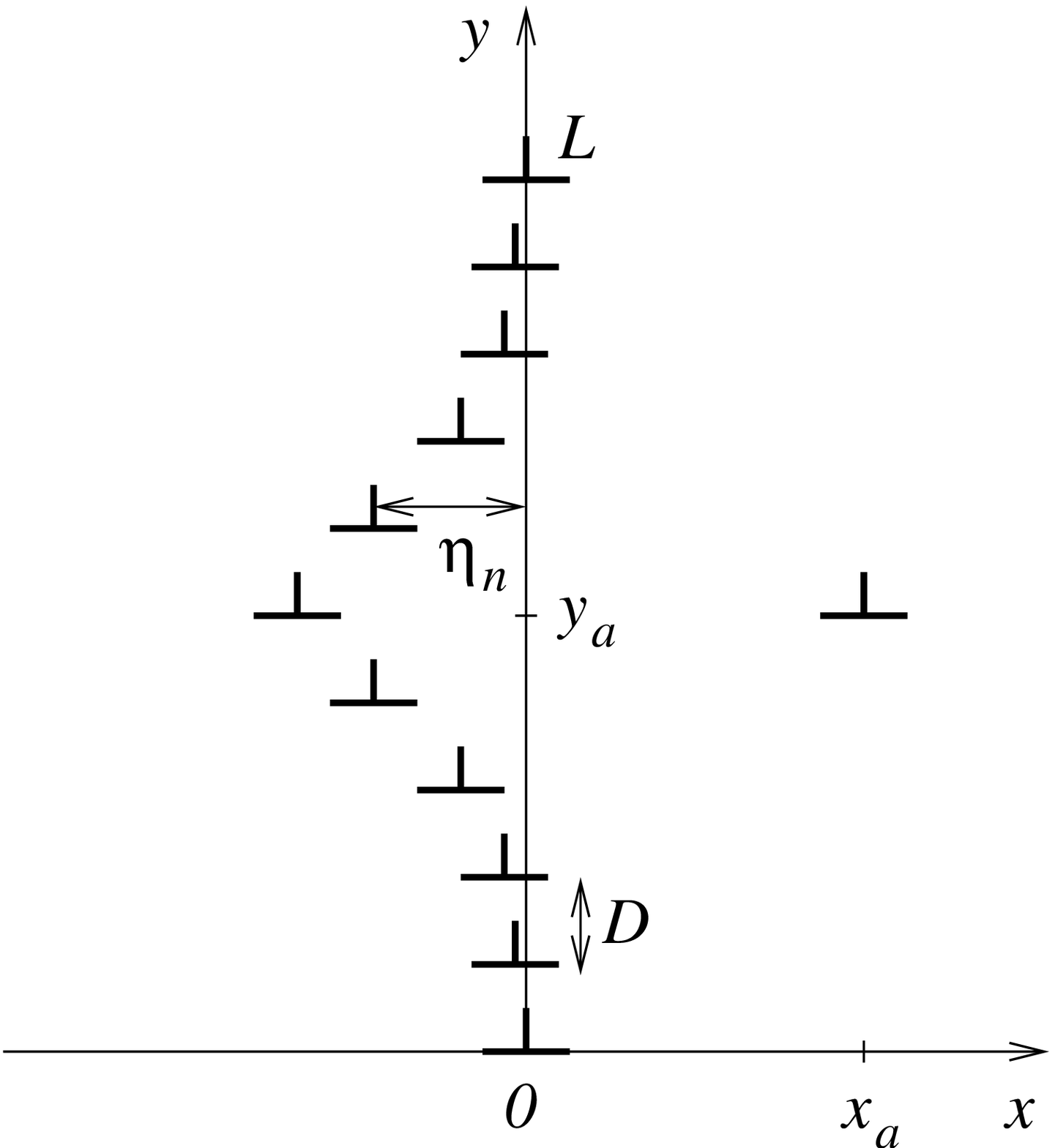}
\end{center}
\caption{A schematic diagram of the grain boundary dislocation system.
The GB is modelled as a discrete linear assembly of $N+1$ edge dislocations
distributed along the $y$ direction, with Burgers vectors $\vec{b}$ parallel
to the $x$ axis. Dislocation spacing is $D$. The endpoint dislocations of
the GB, positioned at $y=0$ and $y=L$, are fixed to $x=0$. The rest of the
GB dislocations are allowed to move in the $x$ direction as a response to
the stress field of the external dislocation positioned at $x=x_a$, $y=y_a$.
The ensuing GB displacement in the $x$ direction is represented by the
vector $\vec{\eta}$, with components $\eta_n$ corresponding to the displacement
of the $n$th dislocation.}
\label{setup}
\end{figure}

Let us consider a small-angle grain boundary of length $L$ in two
dimensions. The GB is schematized as a discrete linear assembly of
$N+1$ edge dislocations distributed along the $y$ direction, with
Burgers vectors ${\bf b}$ parallel to the $x$ axis, of sign $s=\pm 1$. The dislocation
spacing is assumed constant and set equal to $D$. The GB is pinned
at the boundaries ($y=0$ and $y=L$) and can be deformed under an
applied shear stress, see Fig. \ref{setup}.

By representing the GB displacement through the vector ${\eta}$,
where $\eta_n$ is the displacement of the $n-$th dislocation along
 the $x$ axis, the elastic energy cost of a given deformation ${\eta}$ reads
\begin{equation}\label{energy}
E[{\eta}]=\frac{K}{2}\sum_{m\neq
n}\frac{(\eta_m-\eta_n)^2}{(m-n)^2},
\end{equation}
where $K=\mu b^2/(4\pi D^2(1-\nu))$. In our notations, $\mu$ is the
shear modulus and $\nu$ the Poisson ratio. Sums in Equation
(\ref{energy}) run over all couples of dislocations. As a result of
long range dislocation interactions, Equation (\ref{energy}) has the
form of a non-local elastic functional. Its elastic kernel scales as
$|k|$ in Fourier space, where ${\bf k}$ is the wave vector
associated with the GB deformation profile.

If an external stress field $\tau({\bf r})$ is applied, forces acting on the
GB can be calculated in terms of Peach-Koehler interactions. The GB begins to
deform and settles in a configuration where the applied forces equate the
 restoring elastic tensions deriving form the term [\ref{energy}] above.
 Given the particular geometry of our problem, the GB profile is given by
 the displacement field ${\eta}$ which satisfies
\begin{equation}\label{equilibrium_1}
\frac{\delta E}{\delta \eta_n}=sb\tau({\bf r}),
\end{equation}
where the restoring force on the right-hand side is calculated as
the (variational in the continuum case) derivative of the elastic
energy functional in (\ref{energy}). Carrying on the above
calculation, Equation (\ref{equilibrium_1}) can be rewritten as a
simple linear non-homogeneous problem, in the form
\begin{equation}\label{equilibrium_2}
\sum_mV_{nm}\eta_m=sb\tau({\bf r}),
\end{equation}
by defining
\begin{eqnarray}
V_{nm}=\left\{ \begin{array}{cr}
-K\frac{1}{(n-m)^2}, & n\neq m \\
K\sum_{k\neq n}\frac{1}{(n-k)^2}, & n=m.
\end{array}\right.
\end{eqnarray}

\subsection{A simple case - uniform stress}

Let us first review the simple case in which the  applied stress
is uniform, i.e. $\tau({\bf r})=\tau$ \cite{BRU-82}. In order to
solve Equation (\ref{equilibrium_2}), which can be rewritten in
operatorial terms as
\begin{equation}
VH=sb\tau \ \Leftrightarrow \ H=V^{-1}sb\tau,
\end{equation}
one needs to diagonalize the matrix of interactions $V_{nm}$. This
corresponds to solving the eigenvalue problem
\begin{equation}
\sum_m V_{nm}\gamma^l_m=\lambda_l\gamma^l_n,
\end{equation}
where we have called $\lambda_l$ each of the $N$ eigenvalues and
$\gamma^l_n$ each component of the eigenvector corresponding to
$\lambda_l$. It can be shown that for large enough $N$, the
eigenvalues read
\begin{equation}
\lambda_l=\frac{\pi^2 l}{N}\frac{\mu b^2}{2\pi (1-\nu)D^2}=2K\frac{\pi l}{N}
\propto \frac{1}{L}
\end{equation}
and the eigenvectors
\begin{equation}
\gamma^l_n=\sqrt\frac{2}{N} \sin\frac{\pi ln}{N}, l=(0),1,2,3,...N.
\end{equation}
Such an eigenvector representation is possible since the functions
$\gamma^l_n$ match boundary conditions and constitute a {\it
complete} set, such that any continuous function with piecewise
continuous and differentiable derivative can be expanded in terms of
them \cite{FET}. Orthogonality of eigenvectors is ensured  by the
following condition
\begin{equation}
\frac{2}{N}\sum_{n=0}^{N}\sin\frac{\pi l_1 n}{N} \sin\frac{\pi l_2 n}{N}=\delta_{l_1l_2}.
\end{equation}
Now $V=\Gamma^T\Lambda\Gamma$ and $V^{-1}=\Gamma\Lambda^{-1}\Gamma^T$, where we
have called $\Gamma$ the matrix of eigenvectors and $\Lambda$ the (diagonal)
matrix of eigenvalues and we have exploited the orthogonality of the
transformation $\Gamma$, such that $\Gamma^{-1}=\Gamma^T$.
Easily, we derive $H=b\tau\Gamma^T\Lambda^{-1}\Gamma$ in the form
\begin{equation}
\eta_n=sb\tau \sum_{m,l}\lambda^{-1}_l\gamma^l_n\gamma^l_m
\end{equation}
and replacing the sum over $m$ with an integral (this is done also
in the following - see the Appendix for a discussion)
\begin{equation}
\eta_n=s\frac{8}{\pi^2}\frac{1-\nu}{\mu }\frac{DL}{b}\tau\sum_{l}
\frac{1-(-1)^l}{2l^2}\sin\frac{\pi ln}{N},
\end{equation}
where the $[1-(-1)^l]/2$ factor simply limits the sum to odd values of the $l$ index.

\subsection{Stress generated by a dislocation}

Solving this problem in the case of a more complicated form of the
external stress can be a difficult task, instead. In particular, we
are interested in the shear stress generated by an external edge
dislocation $a$, placed at ${\bf r}_a=(x_a,y_a)$. The dislocation
$a$ has a Burgers vector of modulus $b$ and sign $s_a=\pm 1$. Even
assuming that the position of such a dislocation is fixed, the
magnitude of the GB-dislocation interaction will necessarily depend
on the deformation of the GB itself. Equation (\ref{equilibrium_2})
becomes
\begin{equation}\label{equilibrium_3}
\sum_mV_{nm}\eta_m=C_n({\bf r}_a)\times(\eta_n-x_a)    ,
\end{equation}
with $C_n({\bf
r}_a)=ss_ac\times[x_a^2-(Dn-y_a)^2]/[x_a^2+(Dn-y_a)^2]^2$ and $c=\mu
b^2/[2\pi(1-\nu)]$. Here we have assumed $x_n\ll x_a$, as the
non-local stiffness of the GB would not allow huge deformations to
take place. The equation above can be rewritten in the form of
Equation (\ref{equilibrium_2}) as
\begin{equation}
\sum_m[V_{nm}-C_n({\bf r}_a)]\eta_m=-x_aC_n({\bf r}_a)
\end{equation}
or, equivalently,
\begin{equation}
(V-C)H=-x_aC,
\end{equation}
where $C$ is an operator defined accordingly. The obvious solution
would be
\begin{equation}
H=-x_a(V-C)^{-1}C,
\end{equation}
however, unlike the case of a uniform stress, the inversion of the
$(V-C)^{-1}$ matrix is a tough task. We cannot proceed through a
straightforward diagonalization as before, since the matrices $V$
and $C$ are never simultaneously diagonal. A feasible way to carry
on the calculation requires an approximation. Matrix $C$ includes
corrections coming from the deformation of the GB, which is supposed
to be a relatively small effect compared to the main interaction
carried by $V$. These considerations can be easily reformulated in a
more quantitative fashion. As a consequence, using simple relations
of matrix algebra, we can write
$(V-C)^{-1}=(1-V^{-1}C)^{-1}V^{-1}\approx (1+V^{-1}C)V^{-1}$, which
is a first order expansion in $C$. The displacement is then
\begin{equation}
H=-x_a (\Gamma^T\Lambda^{-1}\Gamma +
\Gamma^T\Lambda^{-1}\Gamma C \Gamma^T \Lambda^{-1} \Gamma) C
\end{equation}
or, in the other notation,
\begin{equation}\label{eta_1}
\eta_n=-x_a\sum_m\left[ \sum_{l}\lambda^{-1}_l\gamma^l_n\gamma^l_m +
\sum_{l}\sum_{s}\lambda^{-1}_l\lambda^{-1}_s\gamma^l_n\gamma^s_kC_k({\bf r}_a)
 \gamma^l_k\gamma^s_m\right]C_m({\bf r}_a).
\end{equation}

In Equation (\ref{eta_1}), the second term within square brackets is considerably
smaller than the first and can be neglected. The deformation of the GB becomes
\begin{equation}\label{eta_2}
\eta_n\approx -x_a \sum_{l}\lambda^{-1}_l\gamma^l_n\sum_m\gamma^l_m C_m({\bf r}_a).
\end{equation}
So far we have made basically two assumptions: i) we deal with a
large GB; ii) the external dislocation is not in the immediate
vicinity of the GB. These choices have allowed us to take continuum
limits in sums and to disregard higher order terms. At this point,
however, the sum over $m$ in Equation (\ref{eta_2}) cannot be
calculated easily, unless other approximations are made. We restrict
ourselves to the limit in which the external dislocation sees a
quasi-infinite GB, i.e. $L\gg x_a$. The forces exerted by the
external dislocation on the GB dislocations can be either attractive
or repulsive depending on the distance and the angle between the
external dislocation and each GB dislocation. As a result the
concavity of the GB profile may change along the $y$ direction. The
sum over $m$ in Equation (\ref{eta_2}) can be replaced by an
integral from $-\infty$ to $+\infty$ (see Appendix). The integral is
solved calculating residues around the poles of the $C_m({\bf r}_a)$
function. The displacement thus determined is
\begin{equation}\label{eta_a}
\eta_n=-2ss_ax_a\frac{1}{N}\sum_l e^{-\frac{\pi l}{L}|x_a|}
\sin\frac{\pi ly_a}{L} \sin\frac{\pi ln}{N}.
\end{equation}

\begin{figure}[!h]
\begin{center}
\includegraphics[width = 10cm]{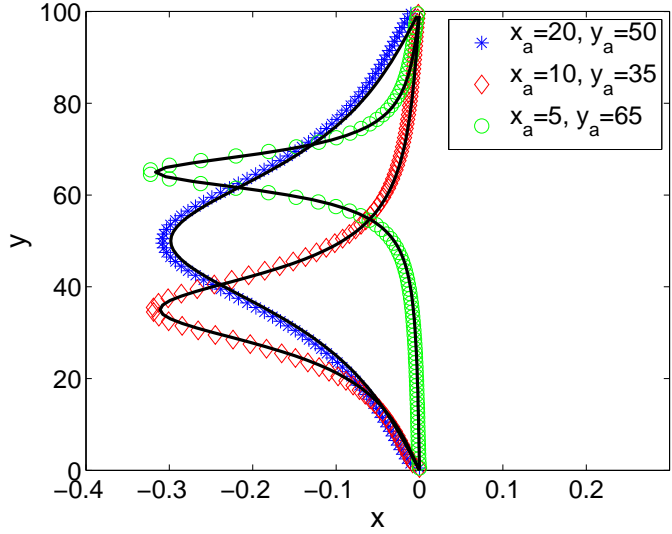}
\end{center}
\caption{Three examples of deformed grain boundary profiles (symbols) in the case
  $L=100 \gg x_a$. The deformation
  is due to the stress field of an external dislocation positioned at $x=x_a$, $y=y_a$.
  The solid lines correspond to the predictions of Eq. (\ref{eta_a}). 500 pinned
  extra dislocations (with the same spacing as in the deformable part) have been
  inserted on both sides of the GB. Notice that there are no fitting parameters.}
\label{gb_profiles}
\end{figure}

This result is compared with numerical results from discrete
dislocation dynamics simulations of the GB-dislocation system.
These are done with a similar setup as e.g. in Ref. \cite{MIG-02},
with a straight GB as an initial configurations. By letting
the GB relax in the stress field of an external dislocation,
a good agreement is found (see Fig. \ref{gb_profiles}). Notice 
that in order for the simulation setup to correspond to Eq. 
(\ref{eta_a}) (which was derived by computing the integral  from 
$-\infty$ to $+\infty$), we have considered a system where in 
addition to the deformable part of the GB of length $L$, there is 
a large number (here 500) of pinned dislocations on both sides 
of the deformable part of the GB. The agreement with Eq. 
(\ref{eta_a}) is the better the larger the number of these 
pinned dislocations.

\subsection{Correction to the GB field}

We are finally able to determine how a dislocation induced
deformation affects the stress field generated by the GB. We
consider the case in which $L\gg x_a$, since it matches are
simulation setups. The shear stress at position $(x,y)$ due to the
deformed GB is a functional of the displacement field $\eta_n$ and
can be expressed, by calling $\sigma_n$ the shear stress of the
$n-$th GB dislocation, as
\begin{equation}\label{stress_functional}
\sigma(x,y)=\sum_n\sigma_n(x-\eta_n,y-y_n),
\end{equation}
where $\eta_n$ is given by Equation [\ref{eta_a}]. Since we are
interested in plastic flow activation well inside each grain, we shall
assume $\eta_n\ll x$, that is we keep far enough from the stress
singularities of GB dislocations. This way we can express stress fields
through their Taylor expansion around $\eta_n\approx 0$ as
$\sigma_n(x-\eta_n,y-y_n)\approx\sigma_n(x,y-y_n)+
\sigma_n'(x,y-y_n)\eta_n+1/2\sigma_n''(x,y-y_n)\eta_n^2+\mathcal{O}(\eta_n^3)$.
Equation (\ref{stress_functional}) will be rewritten as
\begin{equation}\label{stress_sum}
\sigma(x,y)=\sigma^{(0)}(x,y)+\sigma^{(1)}(x,y)+\sigma^{(2)}(x,y)+...
\end{equation}
where the $i$-th term on the right hand side is obtained summing the $i$-th
order as in Equation (\ref{stress_functional}). In passing we note that,
given the nature of the derivation operation, in Equation (\ref{stress_sum})
even terms will be odd functions of $x$, while odd terms will be even
functions of $x$. Then it is no coincidence that $\sigma^{(0)}(x,y)$ is the
well known stress field generated by a straight GB (see References
\cite{HIR,LAN}).

\begin{figure}[!h]
\begin{center}
\includegraphics[width = 4.4cm]{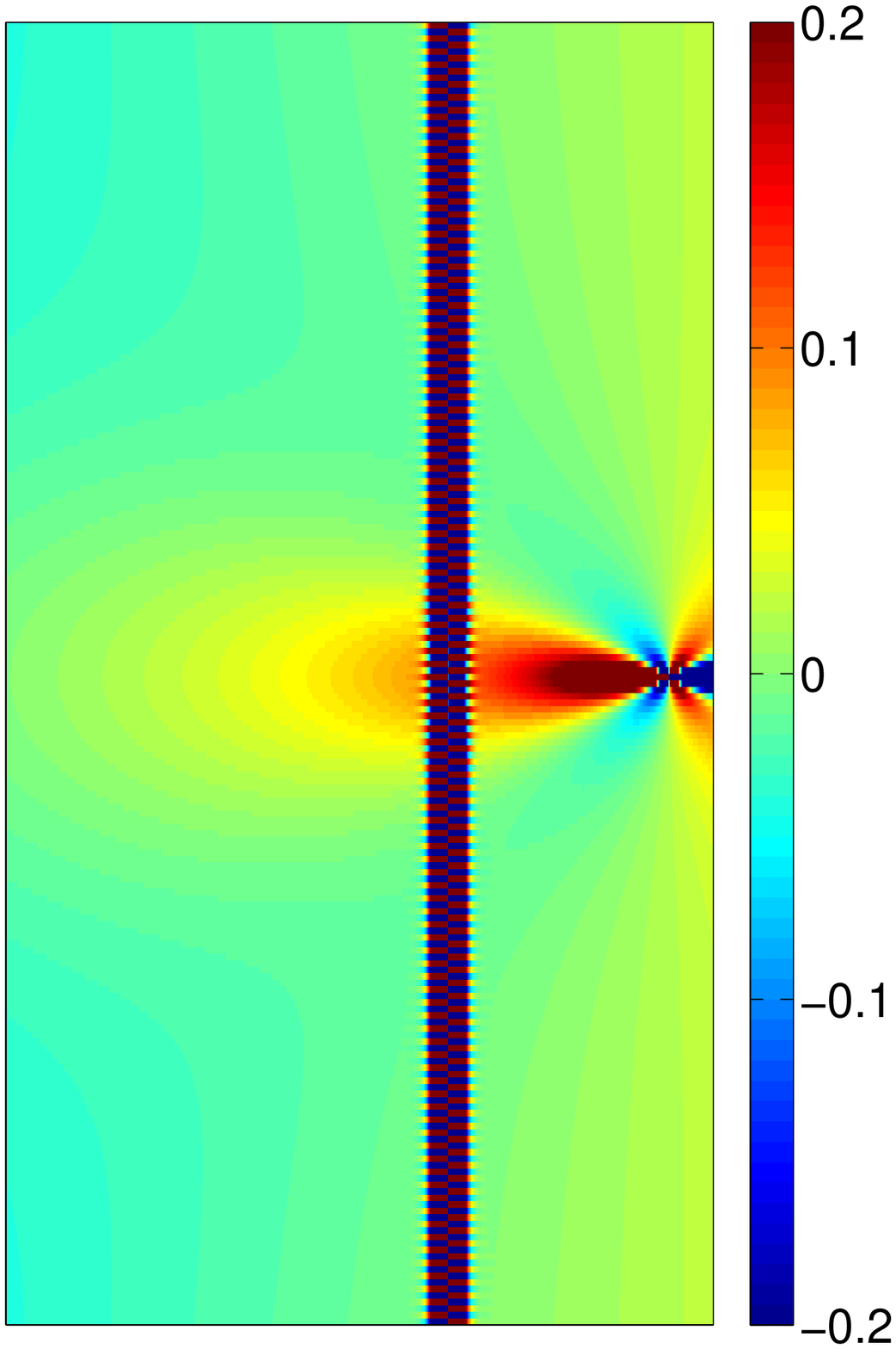}
\includegraphics[width = 4.4cm]{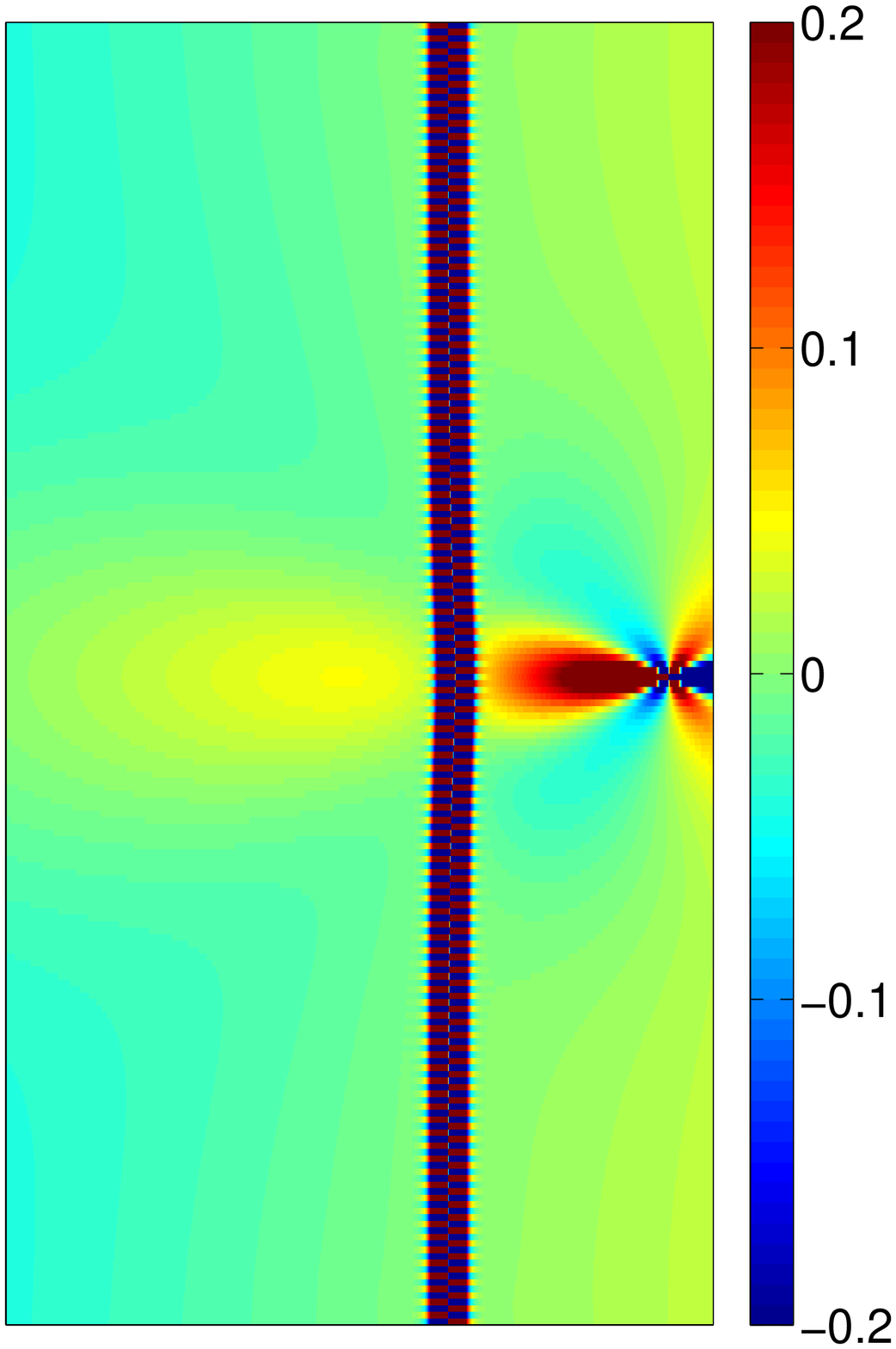}
\includegraphics[width = 4.4cm]{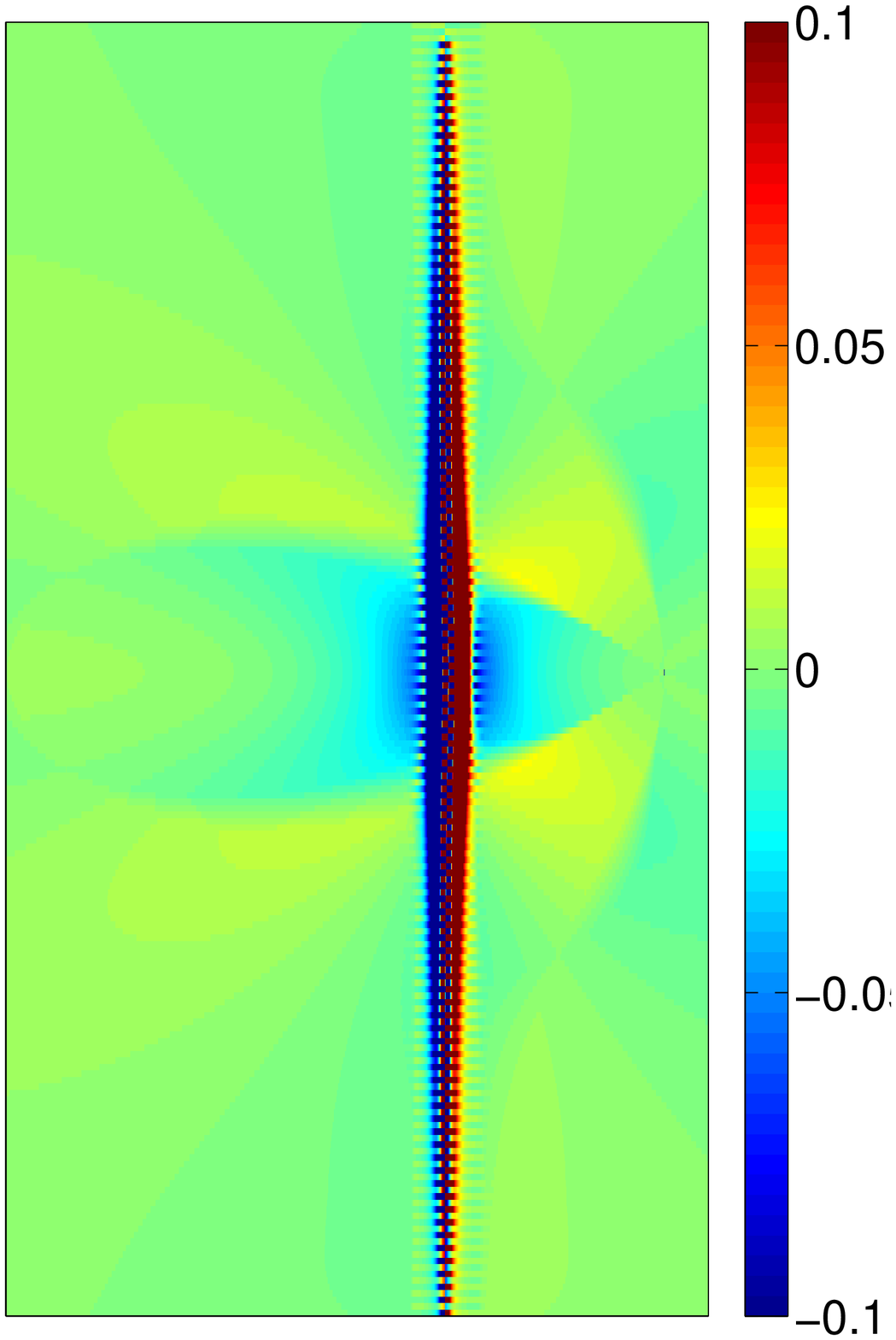}
\end{center}
\caption{
  Stress fields of a system consisting of a grain boundary (the dark vertical
  structure) and an external dislocation positioned to the right side of the
  grain boundary ($x_a=10$, $y_a=50$, $L=100$).
  {\bf Left}:
  Stress field in a situation where the grain boundary is not allowed to deform.
  {\bf Middle}: Stress field with a deformable grain boundary. {\bf Right}: The
  difference of the absolute values of the two stress fields,
  $|\sigma^{deformed}|-|\sigma^{undeformed}|$. Negative values indicate
  screening, positive values enhancement of the magnitude of the stress due to
  grain boundary deformation.}
\label{stressfields}
\end{figure}

As for the first order $\sigma^{(1)}(x,y)$ calculations are not different
from the ones carried on in the previous sections. We eventually obtain,
in the continuum limit of the sum,
\begin{equation}\label{stress}
\sigma^{(1)}(x,y)=\pi s_ax_a\frac{\mu b}{1-\nu}\frac{1}{L^2}\sum_l e^{-\frac{\pi l|x_a|}{L}}
\sin\frac{\pi ly_a}{L}\times
\end{equation}
$$
\times \, l\left(1-\frac{\pi l |x|}{L}\right)e^{-\frac{\pi l|x|}{L}}\sin\frac{\pi ly}{L}.
$$
One can observe that this term is, as expected, an even function of
$x$. It accounts for corrections observed along the $x$-direction,
where the GB deformation produces an even decrease in the absolute
value of the shear stress, see Fig. \ref{stressfields} for corresponding
results from dislocation dynamics simulations. This is balanced by 
an increase along the $45$-degree directions along which shear 
stress is redistributed between adjacent grains. Such corrections 
are introduced by the $\sigma^{(2)}(x,y)$ term, which can be 
calculated integrating the second-order term of the expansion of 
$\sigma_n(x-\eta_n,y-y_n)$. The procedure does not differ from what 
we have seen so far, with the exception of some additional 
complications arising from the fact that this time the sum over 
$l$ is squared. Skipping the details of the calculation, we obtain
\begin{equation}\label{stress_2nd}
\sigma^{(2)}(x,y)=2s\pi^2x_a^2\frac{\mu b}{1-\nu}\frac{D}{L^4}\sum_l\sum_{\lambda}
e^{-\frac{\pi (l+\lambda)|x_a|}{L}}\sin\frac{\pi ly_a}{L}\sin\frac{\pi \lambda y_a}{L}\times
\end{equation}
$$
\times \, \mbox{sign}(x)\left[
(l+\lambda)^2 e^{-\frac{\pi |l+\lambda||x|}{L}}\left(\frac{\pi}{2}
\frac{|l+\lambda|}{L}|x|-1\right)\cos\frac{\pi (l+\lambda)y}{L}-\right.
$$
$$\left.-(l-\lambda)^2 e^{-\frac{\pi |l-\lambda||x|}{L}}\left(\frac{\pi}{2}
\frac{|l-\lambda|}{L}|x|-1\right)\cos\frac{\pi (l-\lambda)y}{L}
\right],
$$
where, as usual, $l,\lambda=1,2,3...N$. Indeed, this term proves to
be an odd function of the variable $x$. The same method allows the
calculation of further orders of the stress expansion
(\ref{stress_sum}) which become relevant as test dislocations are
close to the GB, where the anisotropy of stress fields becomes
relevant.

We should point out that the terms of the sums over $l$ (and $\lambda$) are
exponentially suppressed. As a consequence, sums can be stopped
at the first terms. In particular, if the asymmetry or the {\it
exact} shape of the GB profile are not a concern, the very first
terms provide a fair description of the deformed GB.
Furthermore, one should notice that the correction itself decays
exponentially with the distance of the external dislocation $a$.
This suggests the emergence of a {\it screening} distance, beyond
which the effects of GB deformation do not affect dislocation-GB
interaction anymore. This distance is expected to be proportional
to $L/(\pi l)$, i.e. higher order corrections in $l$ are weaker
at large distance, but may come into play on a shorter range.

\section{Discussion}\label{discussion}
Both analytical calculations and numerical studies of the
stress fields of a GB-dislocation system show that the GB deformation
tends to {\it screen} the dominating stress that is causing the
deformation. At the same time, due to the fact that a stress field
of an edge dislocation has both positive and negative regions, some
parts of such a stress field are also {\it enhanced} in the same
process. One should note, however, that the strongest effect the GB
deformation has is screening close to the deformed part of the GB.
This happens on both sides of the GB (see Fig. \ref{stressfields}).

The implications for collective dynamics in polycrystals with
flexible grain boundaries would then be the following: Low angle
GB's tend to accommodate the excess stresses within a grain by
deforming in such a way that such excess stresses are partly
screened. This screening effect is exponentially suppressed beyond a
length that is of the order of the typical grain size $\sim L$,
and has a magnitude scaling as $\sim 1/L$. Consequently, in addition 
to acting as boundaries to the dislocations themselves, deformable 
GB's tend to hinder also stresses from being transferred from one 
grain to another. This effect gets stronger with decreasing grain
size. One should note, however, that also various dislocation 
arrangements present within grains tend to screen the long range 
stress fields typical to these systems.

The behavior of this simple model system is hence controlled 
mainly by a single parameter, namely the typical grain size 
$\sim L$. In fact, our results suggest that deformable low angle 
GB's tend to confine avalanches of plastic activity to some 
extent within single grains. This could be anticipated to be the 
case also for general (high angle) grain boundaries: even if 
they cannot be thought as deformable arrays of mobile 
dislocations, the large difference in the respective orientations 
of neighboring grains means that only a fraction of the stress 
is effectively transferred over such a GB. 

Recent experiments on polycrystalline ice reveal an intriguing
crossover in the AE amplitude distributions: Small and large
avalanches appear to be characterized by different power law
exponents \cite{RIC-05}. The observation that small avalanches are
described by the same exponent as avalanches in single crystal
samples leads to the natural conclusion that such small avalanches
do not feel the presence of the GB's. On the other hand, larger
avalanches will necessarily interact with the GB's. In the
experiment presented in Ref. \cite{RIC-05}, the exponent describing
such large avalanches was found to be smaller as compared to the
one describing the statistics of smaller avalanches. 

Thus, the presence of GB's will change the statistics of avalanches
in some way. One possibility, along similar ideas as presented in 
Ref.~\cite{RIC-05}, is to consider the spreading process on a 
coarse-grained scale of grains. An initial avalanche within a grain,
if reaching the region close enough to a grain boundary, may spread
to neighboring grains thus triggering avalanches there. Triggering is 
governed by an effective spreading probability
$p$ that depends (among other things) on the details of the 
dislocation-grain boundary interaction discussed above. Hence the ensuing
spreading process is characterized by an average $p$ and geometrical 
factors (correlations in GB orientations, grain sizes and their
correlations etc.). This leads to a number $N$ of grains that
participate in the ``total'' avalanche. The statistics of $N$ will follow at the
simplest a mean-field branching process for which one in general has 
three possible scenarios: {\it i}) a sub-critical process with 
$\langle N \rangle = \mathrm{const}$, {\it ii}) an exactly critical process 
with $P(N) \sim N^{-3/2}$, {\it iii}) a supercritical process which never 
stops. Clearly, the last case can be excluded by common sense. 
As for the the first two options, it appears much more likely that 
the grain-to-grain spreading is actually sub-critical. This would
be in agreement with the observation that in 
experiments of Ref.~\cite{RIC-05} a grain-size dependent cut-off for the
acoustic emission amplitudes was observed. On the other hand,
this would indicate that the observed avalanche distributions would 
not change from what one sees on a single grain level. 

In conclusion, we have considered the GB mediated dislocation
interactions for a ``toy" system in two dimensions. This has allowed
us to make analytical progress, to compare with numerics, and to
understand the nature of screening. Regardless of the obvious limitations
of the simple model studied here, working out these problems leads 
to an increased understanding of the physics of avalanches 
in the plasticity of polycrystalline materials. Clearly, there is much 
room for short-term investigations in three separate directions at 
least. One is the accumulation of further experimental data on these 
phenomena. The second is the influence of the geometry (dimension, 
GB angle, grain size, correlations) on these processes, which can 
be studied partly experimentally and partly by numerical, atomistic, 
simulations. Finally, there seems to be room for the development of 
coarse-grained models which would match their predictions with 
experimental data.

{\it Acknowledgements -} The authors would like to acknowledge the
support of the Center of Excellence -program of the Academy of
Finland, and the financial support of the European Commissions NEST
Pathfinder programme TRIGS under contract NEST-2005-PATH-COM-043386.
MJA and LL would like to thank the ISI institute (Turin, Italy) for
hospitality. PM is grateful to the Laboratory of Physics, Helsinki University of
Technology (Finland) for kind hospitality and acknowledges financial support from
the Ministerio de Educaci\'on y Ciencia (Spain), under grant 
FIS2007-66485-C02-02.

\section*{Appendix}
In several cases in the derivations, sums have been replaced by
integrals to infinity. While the validity of the {\it infinity}
assumption is quite intuitive if $L$ is taken significantly larger
than $x_a$, the transition to continuum requires a few words of
discussion. Provided that the sum is extended to infinity, we
calculate it using the Poisson integral relation (see \cite{LAN})
\begin{equation}
\sum_{n=-\infty}^{\infty}f(n)=
\sum_{k=-\infty}^{\infty}\int_{-\infty}^{\infty}\!dn\,f(n)e^{2i\pi kn}.
\end{equation}
Calculating the integral first, we always obtain an exponential
$\approx e^{i\pi kx/D}$ term, which rapidly goes to zero if $k$ is
increased, as we always assumed $x\gg D$. As a result, the $k=0$
term of the sum over $k$ alone provides a good estimate of the
initial sum. But this is equivalent, as it can be easily noticed,
to replacing the sum with an integral in the first place.

\end{document}